# Enhancing Autonomous Vehicle Design and Testing: A Comprehensive Review of AR and VR Integration

Emanuella Ejichukwu[1], Lauren Tong[1], Gadir Hazime[1], and Bochen Jia[1]

[1]Department of Industrial & Systems Engineering, University of Michigan-Dearborn, MI, USA.

## Abstract

This comprehensive literature review explores the potential of Augmented Reality (AR) and Virtual Reality (VR) technologies to enhance the design and testing of autonomous vehicles. By analyzing existing research, the review aims to identify how AR and VR can be leveraged to improve various aspects of autonomous vehicle development, including: creating more realistic and comprehensive testing environments, facilitating the design of user-centered interfaces, and safely evaluating driver behavior in complex scenarios. Ultimately, the review highlights AR and VR utilization as a key driver in the development of adaptable testing environments, fostering more dependable autonomous vehicle technology, and ultimately propelling significant advancements within the field.

## 1. Introduction

Self-driving vehicles have a much longer history than many may believe with initial testing beginning in the mid-1900s (Scurt et al., 2021). A significant milestone was reached in the 1980s when Carnegie Mellon University's NavLab robotics team developed one of the earliest self-driving cars, leveraging image processing algorithms and laser sensing to navigate predetermined routes (Stanchev & Geske, 2016). Since then, automotive manufacturers have consistently pushed the limits of this technology, leading to improved accessibility of autonomous vehicles, and a seamless integration into the public streets of several major cities (Bimbraw, 2024). However, a fundamental challenge persists, understanding and predicting human reactions to autonomous vehicles (Guanetti et al., 2018). This challenge is paramount as misconceptions about human responses can hinder acceptance and safe incorporation of autonomous vehicles. Traditional testing methods, while invaluable, often fall short in comprehensively assessing the diverse range of human emotions and behaviors encountered in real driving scenarios. This knowledge gap risks discrepancies between AV programming and actual driver responses, potentially compromising safety and eroding public trust in autonomous vehicle technology.

In response to these challenges, the use of AR and VR has emerged as a promising solution for AV testing and development (Zhang et al., 2020). These technologies offer a controlled environment where realistic, yet safe, driving interactions can be simulated. Researchers can use AR and VR to systematically study and understand a spectrum of human reactions to AVs—ranging from decision-making and emotional responses to situational awareness and compliance with AV operations. This detailed insight is essential for refining AV systems to be more intuitive and aligned with typical human behaviors, thereby enhancing their reliability, safety, and public acceptance. Through the strategic application of AR and VR in testing, the goal is to bridge the gap between human drivers and autonomous technologies, fostering a safer and more harmonious integration of AVs into public roadways.

Recent literature has shown a growing interest in utilizing AR and VR technology for testing autonomous vehicles. This experimentation has garnered considerable positive responses (Kettle & Lee, 2022), leading to expanded use of virtual AR and VR driving simulations to assess user experiences (Hamad & Jia, 2022; Postelnicu & Boboc, 2024). Additionally, advancements are being made in comprehending driver behaviors and customer experiences (Lemon & Verhoef, 2016; Xu et al., 2022).

Virtual reality transports users to a simulated reality, creating a completely immersive experience where they can interact with digital surroundings as if they were physically present (Boboc et al., 2020). It can replicate the real world or create entirely fictional environments. This is achieved using computer technology and VR headsets that provide users with multi-dimensional images and environments, allowing them to look around, move within, and interact with the virtual spaces as if they were physically inside them (Riegler et al., 2020). VR creates a fully immersive experience where the user's vision is completely taken over by the display (Rehman et al., 2018).

Augmented reality, on the other hand, enhances the real-world environment by overlaying computer-generated graphics onto the natural surroundings, effectively augmenting the user's experience (Gremmelmaier et al., 2022). It layers digital enhancements atop an existing reality or live view, often via smartphones, tablets, or AR glasses. Augmented reality can improve our knowledge of driver behavior by projecting information onto the windshield or other convenient surfaces, helping to guide driver actions and decisions (Bran et al., 2020). AR can alert drivers of hazards and give feedback on driver performance, which can be used for data collection and real-time coaching (Ng-Thow-Hing et al., 2013; Uchida et al., 2017). It adds graphics, sounds, and haptic feedback to the natural world as it exists. In contrast with VR, augmented reality enhances one's current perception of reality, whereas virtual reality replaces the real world with a simulated one (Hu et al., 2023).

This literature review underscores the critical importance of understanding driver behavior in the development and testing of autonomous vehicles. By examining existing research, this review aims to explore how AR and VR technologies can inform AV design and testing methodologies. AR and VR offer novel platforms for researchers to model realistic driving events in controlled environments (Fu et al., 2013; Lee et al., 2015). This capability allows drivers to interact with AVs without the inherent hazards associated with field testing. This is particularly valuable for studying driver reactions in complex, risky, or unfeasible real-world scenarios (Bagloee et al., 2016; Kurt et al., 2014). The ability to create precisely controlled environments with quantifiable variables makes AR and VR ideal tools for exploring human-machine interactions and driver decision-making processes in greater depth (Hiroto Suto et al., 2020; Sun et al., 2019). This level of control allows researchers to gather detailed data on user responses to various scenarios.

This review critically analyzes the advantages, limitations, and potential applications of AR and VR technologies in understanding driver behavior during AV testing. It explores the current landscape of AR and VR use in AV development and proposes promising avenues for future research. By leveraging these technologies, researchers can contribute to the development of more reliable, user-friendly AV systems, ultimately fostering the creation of more human-centric driverless vehicles.

## 2. SCOPING LITERATURE REVIEW

### 2.1 Search Strategy

For this review, academic databases covering a broad spectrum of VR and AR research, such as ProQuest, Elsevier, Research Gate, IEEE Xplore, and Scopus were extensively searched to gather studies related to the topic. Google Scholar, university library, and other digital repositories were searched for journal publications, research reports, theses, and dissertations on VR simulations used in automated driving vehicles. The search terms used include virtual reality (VR), virtual reality simulations for autonomous driving, AR in autonomous vehicle testing, VR in autonomous vehicle testing, simulating driver behaviors in AV, and AR/VR training for autonomous vehicle operation.

## 2.2 Research Questions

Due to the present limits of simulation accuracy, investigating developments in sensory feedback, and incorporating complex behavioral models are needed in AV. To better understand driver behavior in autonomous vehicle (AV) environments, this literature review investigates the use of augmented reality (AR) and virtual reality (VR) technologies to provide more lifelike simulations for in-depth analysis of driver behavior during. To conduct the literature review question asked is:

- How have AR and VR technologies been leveraged to gain insights into human behavior within AV?
- How can AR and VR technologies enhance AV design and testing?

Analyzing the potential of AR and VR simulations to mimic real-world behaviors can capture the unpredictable and dynamic nature of driving during AV testing. In turn, this will lead to a deeper understanding of driver decision-making, responses, and trust in autonomous vehicles. Ultimately, this knowledge will contribute to the development of safer, more user-friendly, and widely accepted autonomous driving technologies.

## 2.3 Selection Criteria

The selected papers describe the results of experiments conducted in virtual and augmented reality testing of automated vehicles. To ensure that studies considered remain current, only articles published between 2014 and 2024 were chosen. This range guarantees that the review covers the most recent use of virtual and augmented reality in automated vehicles.

Primary research, empirical studies, case studies, and systematic reviews are some of the study types used to examine the influence, use, and efficacy of VR/AR in automated driving vehicles. The included studies evaluate or examine the technical aspects of using virtual or augmented reality in automated driving vehicles. To ensure effective comprehension and language uniformity, only publications written in English were accepted. The flowchart of the selection process is presented below.

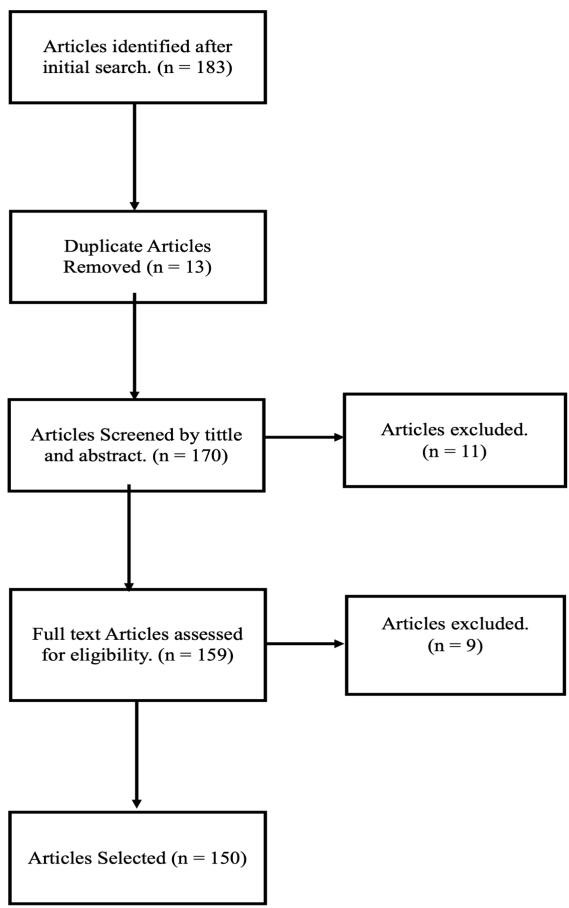

*Figure 1: Flow Chart for Article Selection Process*

### 2.3 Data Extraction

The titles and abstracts of the papers were first gathered to determine their alignment with the review. All studies that did not meet the inclusion criteria or study objectives were screened out. Following this, 150 papers were left. The full texts of the remaining studies were assessed and evaluated based on the inclusion and exclusion criteria to establish their suitability with our proposed questions. Throughout the gathering, arranging, and maintaining references and citations, Zotero was used as a reference manager for the reference list. It was also used for searching, retrieving, and inserting citations and references. Further search will be conducted for more papers.

### 2.4 Quality Assessment

The inclusion, exclusion, and data extraction criteria were specified for articles across all databases. The methodology used in each study was thoroughly reviewed in this evaluation by examining the standing of the authors, the study design, what metrics were used, and how the data collection and analysis processes were carried out. The findings were also further examined. Each study's approach was evaluated for both present and future applicability.

## 2.5 Synthesis of Literature

This phase involved integrating, contrasting, and comparing the findings and insights from many sources. The studies were first arranged chronologically to understand the progress made around the review questions. The areas of agreement and disagreement among studies were also gathered into points, assessing the advantages and disadvantages of the methods presented or previously applied in the studies.

## 3. INTEGRATION OF AR AND VR IN AUTONOMOUS VEHICLE TESTING

This section provides an overview of the general applications of AR and VR in AV testing, such as simulating driving scenarios, studying human factors to improve safety and fostering trust, evaluating driver responses, and training AI systems. It explores the need for AR and VR in the advancement of AV testing, with a particular focus on improving and comprehending driver behavior. By organizing our research around these fundamental uses, this literature review sheds light on how AR and VR technologies work together to better understand driver behavior and create safer and more dependable AV driving experiences.

## 3.1 Driving Conditions Simulation Using VR and AR

AR and VR technologies are being utilized to simulate a wide range of driving conditions, including urban, suburban, and rural environments, as well as adverse weather conditions such as heavy rain, snow, and fog (Kurt et al., 2014). These simulations enable comprehensive testing of automated vehicle systems in diverse scenarios, contributing to the refinement and validation of their performance across different environments (Lee et al., 2016). Drivers can be subjected to numerous driving conditions and their reactions can be studied in detail (Barnard et al., 2016; Khan & Lee, 2019; Morales-Alvarez et al., 2020). By monitoring physiological responses, eye movement, and decision-making processes, it becomes possible to identify behaviors that lead to accidents or near-miss incidents and understand the effect of various factors on driving performance (Darzi et al., 2018; Halin et al., 2021; Lin et al., 2014).

## 3.2 Improving Analysis of Human-Vehicle Interaction

*B*eyond environmental modeling, AR and VR integration in AV testing allows for a deeper examination of the relationship between automated systems and human drivers (Yao et al., 2018). By leveraging AR and VR, researchers can create dynamic and interactive simulations that replicate real-world driving experiences, allowing for the exploration of human factors such as attention, decision-making, and situational awareness in the presence of automated vehicles (Feng et al., 2018). This involves creating virtual scenarios that incorporate these human elements, evaluating how drivers view and react to AV's, and how these vehicles might be designed to detect and adjust to human behavior (Uchida et al., 2017). Participants' psychological responses in VR environments closely mirror their real-world experiences making this data trustworthy and reliable (Slater et al., 2006). These kinds of evaluations are essential for creating AV characteristics and user interfaces that are in line with human instincts and inclinations, which will eventually lead to a mutually beneficial connection between drivers and AVs (Zhang et al., 2020).

## 3.3   Enhancing Perception and Sensor Systems

To test and improve the sensor and perception systems of AV, AR and VR technologies are essential for the visualization of sensor data in a 3D virtual environment, enabling engineers to evaluate the accuracy and robustness of sensors such as LiDAR, radar, and cameras (Nascimento et al., 2019). Researchers can identify and correct such problems by using virtual simulations to assess the efficacy of perception algorithms and the accuracy of sensors in real-time (Ariansyah et al., 2018; Morales-Alvarez et al., 2020). This component guarantees that AV systems can reliably read and navigate the real world. Furthermore, AR and VR facilitate the identification of potential sensor limitations and aid in optimizing sensor fusion algorithms (Gao et al., 2020). Enhancing perception and sensor systems through AR and VR serves as a powerful tool to simulate and analyze complex scenarios without real-world risks. These technologies can help researchers and engineers design better vehicles and systems that consider human driver behavior, ultimately leading to safer and more efficient transportation.

## 3.4   Using AR and VR to Promote Driver Safety and Trust in AV

Enhancing safety and fostering trust are important topics for AVs among end-users and stakeholders. Human-centric design principles guide the creation of immersive simulations that not only replicate driving environments (Gruyer et al., 2017; Kuutti et al., 2018; Rosique et al., 2019). AR and VR allow AVs to be safely tested in dangerous environments without the associated real-world risks with physical testing. They make it possible to create simulations that can boost user and stakeholder confidence in a safe integration onto public roads (Sportillo et al., 2019). By integrating human factors such as perception, cognition, and emotion into the design of VR and AR simulations, researchers create user-friendly systems, and evaluate safety-critical features and emergency response systems in a controlled and repeatable manner, thereby enhancing overall road safety and reliability of automated vehicles (Kurt et al., 2014; Tran et al., 2021).

The integration of VR and AR with simulation platforms can facilitate the assessment of potential safety hazards that may arise from complex driving situations, model actual driving conditions, examine human-vehicle interactions, refine sensor systems, and foster safety and trust thereby contributing to the development of robust safety measures in automated vehicles (Kouatli, 2015). By further exploring these applications and embracing the potential advancements in VR and AR technologies, the industry is positioned to accelerate the development and deployment of AV while simultaneously prioritizing public safety and acceptance (Helgath et al., 2018). This is especially pivotal as self-driving technology continues to evolve, requiring a profound understanding of how drivers will adapt to and coexist with automated driving systems.

# 4 ENHANCING AUTOMATED DRIVING WITH USER BEHAVIOR INSIGHTS

VR and AR have played pivotal roles in understanding driver behavior and enhancing safety, especially in the context of evolving autonomous vehicle technology. This section paints a comprehensive picture of driver behavior in response to AR and VR technologies in the context of automated vehicles, studying it from initial reactions to in-depth analysis of decisions and performance in complex, interactive environments.

## 4.1 User Experience and Acceptance Studies

Understanding how individuals perceive and interact with AR and VR simulations is essential for the successful implementation of these technologies in automated vehicle testing (Fridman, 2018; Lungaro et al., 2018;

Yeo et al., 2020). User experience and acceptance studies delve into the psychological and behavioral aspects of human engagement with virtual environments (Konkol et al., 2020; Lawson et al., 2016). By analyzing user feedback, emotional responses, and trust perceptions, researchers can gauge the effectiveness of AR and VR in conveying the capabilities and limitations of automated driving systems (Babiker et al., 2019). Insights from these studies inform the iterative refinement of VR and AR simulations to align with user expectations and foster trust in automated vehicle technology.

### 4.2 Collaborative Testing and Training Environments

Virtual simulations can provide valuable hands-on experience in testing and validation procedures, contributing to the continuous refinement and improvement of safety protocols for automated vehicles (Schumann et al., 2015; Schwarting et al., 2018). VR and AR facilitate collaborative testing and training scenarios that involve multiple stakeholders, including engineers, regulators, and end-users (Bundell, 2023; Ciprian Firu et al., 2021). By creating shared virtual spaces, AR and VR enable participants to collectively assess the performance of automated vehicle systems, simulate emergencies, and engage in scenario-based training exercises (Carballo et al., 2019; Sportillo et al., 2019). These collaborative environments not only enhance cross-disciplinary communication and decision-making but also contribute to building consensus and trust in the development and deployment of automated driving technologies.

### 4.3 Understanding Driver Behavior with VR and AR:

Researchers are leveraging Virtual Reality (VR) and Augmented Reality (AR) to gain a deeper understanding of driver behavior, which includes the study of their reactions, attention, decision-making, and how they adapt to various driving conditions (Halabi et al., 2017; Sun et al., 2019). VR provides a simulated environment that can replicate the intricacies of real-world driving scenarios, including interactions with AV. This capability is valuable for observing drivers in controlled settings that pose no actual risk, allowing for a comprehensive analysis of their performance. Through the use of eye-tracking and biometric monitoring technologies, involving both VR and AR can be used to monitor driver engagement as well as how attention fluctuates between manual control and autonomous vehicle features (Ahlström et al., 2021; Rangesh et al., 2018).

Additionally, these technologies are instrumental in Human-Machine Interface (HMI) testing by simulating in-car interfaces, thereby enabling researchers to evaluate how drivers interact with and respond to autonomous systems (Naujoks et al., 2019). Such feedback is indispensable for the development of intuitive HMIs that facilitate natural and comfortable interaction between the driver and the vehicle (Detjen et al., 2021). VR also offers the unique advantage of scenario planning, where researchers can introduce drivers to extreme or infrequent situations (Sportillo et al., 2018). This helps in understanding potential reactions and decision-making in scenarios where autonomous systems could malfunction or behave unpredictably, providing critical insights into the development of safer and more reliable autonomous driving technology (Paterson & Picardi, 2023).

## 4.4 Enhancing Safety in AV with VR and AR:

Virtual Reality (VR) and Augmented Reality (AR) technologies are proving essential in enhancing the safety of AV through various innovative approaches. VR is making strides in the development of training programs targeted at drivers, equipping them with the knowledge of both the capabilities and limitations of AV (Riegler et al., 2019). Such educational initiatives ensure that drivers are more adept at taking control of the vehicle when the situation demands it.

AR comes into play by providing drivers with visual cues that outline the perception and intentions of the autonomous system (Phan et al., 2016). By visualizing the car's intended path or highlighting detected obstacles, drivers can gain a better understanding of the vehicle's operation, thus fostering trust in the technology (Wang et al., 2023). Moreover, AR serves a critical role in managing driver distractions by presenting crucial safety alerts and information in a manner that is easily comprehensible, helping to keep the driver's focus on the road ahead (Bran et al., 2020).

When it comes to the design and ergonomics of vehicle interiors, VR and AR are invaluable tools that assist designers in creating spaces conducive to the smooth transition from manual to autonomous driving (Lawson et al., 2016; Xie & Chang, 2021). They can evaluate how various design elements impact driver comfort and control, ensuring that the vehicle interior supports both modes of operation without compromise. By integrating such immersive technologies, the development of AV is becoming increasingly centered on providing a safe, comfortable, and intuitive experience for drivers (Riegler et al., 2020).

In summary, the realm of autonomous vehicle technology, VR and AR are employed through various methods to enhance research and development. Controlled experiments within VR provide researchers the means to simulate diverse driving conditions, traffic scenarios, and pedestrian interactions, which are instrumental in studying and understanding driver behavior when engaging with autonomous driving systems (Lawson et al., 2016). Designers similarly benefit from VR and AR by using these platforms for prototype and user experience testing, where they can quickly model and assess new vehicle control interfaces, making iterative adjustments based on driver feedback (Rasouli & Tsotsos, 2020).

Furthermore, VR offers an immersive training environment for drivers, enabling them to become familiar with the interfaces and operational nuances of AV. This training is crucial for ensuring drivers are prepared to intervene effectively if the system encounters limitations or malfunctions (Nascimento et al., 2019). Safety evaluations also take on a new dimension with VR, allowing researchers to replicate critical situations to assess the reliability of communication between AV and their drivers, especially regarding when and how drivers should regain control (Yao et al., 2018).

These methods help in developing a deeper understanding of how drivers will interact with AV and contribute to creating systems that are safer and more in tune with human behavior (Pauzie & Orfila, 2016). This understanding is essential for improving safety features, designing better human-machine interfaces, and ensuring a smoother integration of AV into daily transportation.

## 5. CHALLENGES AND LIMITATIONS

### 5.1 Replicating Complex and Unpredictable Scenarios and Environments

Although the utilization of virtual reality (VR) and augmented reality (AR) in AV testing is beneficial, these methods cannot replace real-world testing entirely. Driver behaviors in response to unexpected events, interactions with other road users, and environmental factors present a level of complexity that may be difficult to fully capture in virtual simulations (Deter et al.,

2021; Feng et al., 2018). The dynamic nature of driving, including real-time decision-making and adaptability to changing conditions, poses a challenge in accurately representing these aspects within AR and VR environments for the study of driver behaviors.

The controlled nature of simulated environments fails to fully imitate the complexity and unpredictability inherent in real-world scenarios, posing challenges for VR/AR to capture subtle environmental cues and respond to unforeseen situations that occur in reality. Additionally, the current state of VR/AR technology may not be able to entirely replicate the comprehensive sensory inputs crucial for AV, potentially resulting in some inaccuracies in the data collected from these human-behavioral studies.

## 5.2 Technical Limitations

While AR and VR offer the potential to study driver behaviors for AV's, technical limitations pose significant challenges to their effective implementation. One such limitation is the difficulty in achieving a high level of realism to accurately simulate a wide range of driving scenarios (Sivak & Schoettle, 2015). The fidelity of virtual environments created by AR and VR technologies may not always align with the complexity and unpredictability of real-world driving conditions. Achieving this level of realism requires significant computational power and advanced modeling techniques.

This mismatch in realism could impact the validity of studying driver behaviors in simulated environments. Furthermore, the user experience and hardware requirements of AR and VR technologies present challenges in studying driver behaviors. Issues such as motion sickness, visual fatigue, and the need for high-performance computing equipment may limit the feasibility and scalability of using AR and VR for comprehensive research on driver behaviors.

## 5.3 Ethical and Privacy Considerations

The integration of AR and VR for studying driver behaviors in the context of AV raises ethical and privacy considerations. In conducting research using these technologies, ensuring the confidentiality and privacy of participants' data and experiences becomes paramount. Additionally, the ethical implications of exposing participants to potentially risky or challenging scenarios within the simulated environments must be carefully addressed. Balancing the need for authentic and meaningful research with the protection of participants' well-being and privacy presents a complex challenge in the use of AR and VR for studying driver behaviors (Kouatli, 2015).

## 6. FUTURE DIRECTIONS

Possible problems with human-AV interaction in the AV industry may be detected and resolved by utilizing AR and VR, hence opening the door for more intuitive and user-friendly AV. To improve the validity and trustworthiness of the results, subsequent research should encompass a greater variety of behaviors, scenarios, and participant demographics (Derakhshan et al., 2022; Gremmelmaier et al., 2022).

### 6.1 Enhanced Realism and Immersion

Recent research in AR and VR integration for automated vehicle testing has focused on enhancing the realism and immersion of virtual environments. Advanced rendering techniques, higher-resolution displays, and more realistic physics simulations contribute to creating virtual worlds that closely mirror the complexities of real-world driving (Rehman et al., 2018). These developments are crucial for enabling comprehensive testing and validation of automated vehicle systems under a wide range of conditions.

### 6.2 Multi-Sensory Feedback and Interaction

Another notable area of research involves the implementation of multi-sensory feedback and interaction in AR and VR environments for automated vehicle testing (Valášková et al., 2022). By incorporating haptic feedback, auditory cues, and interactive interfaces, researchers are striving to create more engaging and immersive experiences that closely resemble the sensory inputs encountered during actual driving. This multi-sensory approach enhances the fidelity of virtual testing environments and provides valuable insights into human-vehicle interaction dynamics.

## 7. CONCLUSION

VR and AR stand as invaluable tools in comprehending driver behavior and testing complex interactions within safe and controlled environments. They serve as effective mediums for training drivers to interact with autonomous systems confidently, fostering trust and reducing hesitation towards automated technology. By simulating a wide array of scenarios, these technologies provide essential data to enhance AI algorithms governing AV behavior.

Looking ahead, as the field of automated driving evolves, the future of AR and VR in the testing of automated vehicles holds great promise in improving safety, and trust in the use of AV's in public areas. Automated vehicles must be programmed to interpret and predict human driver actions and also communicate effectively with the drivers during transitional control periods (when control shifts between humans and machines). The integration of AR and VR with advanced artificial intelligence and machine learning capabilities is anticipated to lead to the creation of highly adaptive testing environments, which will lead to a more reliable, and deeper understanding of how humans and automated systems can coexist on the roadways, thereby advancing the robustness and reliability of automated vehicle design and technology.